\documentclass[preprint,showpacs,preprintnumbers,amsmath,amssymb]{revtex4}

\usepackage{amssymb}
\usepackage{amsmath}
\usepackage{graphicx}

\usepackage{color}

\begin{document}

\title{Chaoticity parameter $\lambda$ in two-pion interferometry in an expanding boson gas model}

\author{Jie Liu, Peng Ru, Wei-Ning Zhang\footnote{wnzhang@dlut.edu.cn}}
\affiliation{School of physics and optoelectronic technology, Dalian University of Technology,
Dalian, Liaoning 116024, China}
\author{Cheuk-Yin Wong}
\affiliation{Physics Division, Oak Ridge National Laboratory, Oak Ridge, Tennessee 37831, USA}


\begin{abstract}
We investigate the chaoticity parameter $\lambda$ in two-pion interferometry in an expanding
boson gas model.  The degree of Bose-Einstein condensation of identical pions, density
distributions, and Hanbury-Brown-Twiss (HBT) correlation functions are calculated for the
expanding gas within the mean-field description with a harmonic oscillator potential.  The
results indicate that a sources with thousands of identical pions may exhibit a degree of
Bose-Einstein condensation at the temperatures during the hadronic phase in relativistic
heavy-ion collisions.  This finite condensation may decrease the chaoticity parameter
$\lambda$ in the two-pion interferometry measurements at low pion pair momenta, but influence
only slightly the $\lambda$ value at high pion pair momentum.
\end{abstract}
\maketitle

\section{Introduction}

Hanbury-Brown-Twiss (HBT) interferometry is a useful tool to study the space-time structure
of the particle-emitting source, through the two-particle momentum correlation function of
identical bosons \cite{Gyu79,Won94,Wie99,Wei00,Lis05}.  In HBT interferometry, the chaoticity
parameter $\lambda$ is introduced phenomenologically to represent the intercept of the
correlation function at zero relative momentum of the particle pair.  The value of $\lambda$
is related to the degree of source coherence, as it is well known that HBT correlation
disappears for a completely coherent source.

In high energy heavy-ion collisions, pions are the most copiously produced particles.  The use
of  two-pion interferometry to probe the source coherence was proposed at the end of 1970s
\cite{FowWei77,Gyu79}.  However, the experimental measurement of the $\lambda$ quantity is
affected by the presence of many other effects such as  particle misidentification, long-live
resonance decay, final state Coulomb interaction, non-Gaussian source distribution, and so on
\cite{Wie99,Lis05}.  The explanation of the experimental $\lambda$ results is still an open
question.  In 1993, S. Pratt proposed a pion laser model in high energy collisions and studied
the influence of pion laser on two-pion HBT correlation function and the chaoticity parameter
\cite{Pra93}.  In 1998, T. Cs\"{o}rg\H{o} and J. Zim\'{a}nyi investigated the effect of
Bose-Einstein condensation on two-pion interferometry \cite{CsoZim98}.  They utilized Gaussian
formulas describing the space and momentum distributions of a static non-relativistic boson
system, and investigated the influence of the condensation on pion multiplicity distribution.
In 2007, C. Y. Wong and W. N. Zhang studied the dependence of the HBT chaoticity parameter on
the degree of Bose-Einstein condensation for static non-relativistic and relativistic boson
gases within a mean-field with a spherical harmonic oscillator potential \cite{WonZha07},
which can be analytically solved in non-relativistic case and be used in atomic physics
\cite{Pol96,Nar99,Via06}.  The similar work for cylindrical static boson gas sources was just
completed \cite{LiuZha13}.  Recently, the experimental investigation of the source coherence
in the Pb-Pb collisions at $\sqrt{s_{NN}}=2.76$ TeV at the Large Hadron Collider (LHC) was
carried out by the ALICE collaboration \cite{ALICE13}.  A substantial degree of source
coherence was measured \cite{ALICE13}, using a new three-pion interferometry technique with
an improvement over the past efforts \cite{NA44,WA98,STAR}.  It is of interest to study the
reasons of source coherence in heavy-ion collisions at the LHC energy.

In this article we investigate the Bose-Einstein condensation of identical pions for an
expanding relativistic gas source within the time-dependent mean-field of harmonic oscillator
potential, which decreases with time in the temperature region of the hadronic phase, 170--
60 MeV, in relativistic heavy-ion collisions.  Using one- and two-body density matrices, we
calculate the pion space and momentum density distributions, HBT correlation functions, and
the chaoticity parameter $\lambda$ in HBT interferometry in the temperature region of 60 to
170 MeV.  The influences of the pion source Bose-Einstein condensation on the $\lambda$
values in the HBT measurements at different pion pair momenta and temperatures are investigated.
The results indicate that a source with thousands of identical pions may appear to possess a
substantial degree of Bose-Einstein condensation.  The finite condensation may decrease the
chaoticity parameter $\lambda$ in the two-pion interferometry measurements at low pion pair
momenta, but influence only slightly  the $\lambda$ value measured at high pion pair momenta.
In heavy-ion central collisions at the LHC energy, the identical pion multiplicity of event
can reach several thousands.  In this case, the effects of Bose-Einstein condensation on the
chaoticity parameters in two-pion and multi-pion HBT measurements would be of great interest.

This paper is organized as follows.  In Sec. II, we investigate pion Bose-Einstein condensation
for the expanding sources of boson gas in relativistic heavy-ion collision environments.  In Sec.
III, we calculate the pion space and momentum density distributions and two-pion HBT correlation
functions, using one- and two-body density matrices.  We investigate the influences of the
Bose-Einstein condensation on the chaoticity parameter in two-pion interferometry in Sec. IV.
Finally, summary and conclusion are present in Sec. V.

\section{Bose-Einstein condensation in expanding boson gas with harmonic oscillator potential}
\subsection{Time-dependent harmonic oscillator potential}
Base on the previous works of boson gases trapped by a static harmonic oscillator potential in
atomic physics and high energy heavy-ion collisions \cite{Pol96,Nar99,WonZha07,LiuZha13}, we are
interesting in the study of the boson gas of identical pions within a time-dependent harmonic
oscillator potential, $V(\textbf{\emph{r}}, t)$, that arises approximately from the mean-field
of the hadronic medium in high energy heavy-ion collisions \cite{WonZha07}.  The time-dependent
harmonic oscillator potential is given by,
\begin{equation}
\label{Vrt}
V(\textbf{\emph{r}},t) =\frac{1}{2}\,m\,\omega^2(t)\, r^2 =
\frac{1}{2}\,\hbar\,\omega(t)\, \frac{r^{2}}{a^{2}(t)},
\end{equation}
where, $t$ is source evolution time, $m$ is the boson mass, $\hbar \omega(t)$ measures
the time-dependent potential strength, and the characteristic length of harmonic oscillator
$a$ is defined as $a(t)=\sqrt{\hbar/m\omega(t)}$.  We assume that the characteristic length
is proportional to a parameterized source radius which increases with time as $R=R_0+\alpha t$,
where $R_0$ is the initial radius of source and $\alpha$ is a parameter related to the source
average expansion velocity and will be determined by hydrodynamics.  For ideal boson gas, the
system energy is simply the summation of all individual bosons, and the energy levels of a
boson for given $\hbar \omega$ are
\begin{equation}
\label{enen}
\varepsilon_n=n \hbar\,\omega + \frac{3}{2} \hbar\,\omega,
~~~~n=0,1,2,...\,.
\end{equation}
The degeneracy of $\varepsilon_n$ is $g_n=(n+1)(n+2)/2$.

\begin{figure}
\includegraphics[width=0.4\columnwidth]{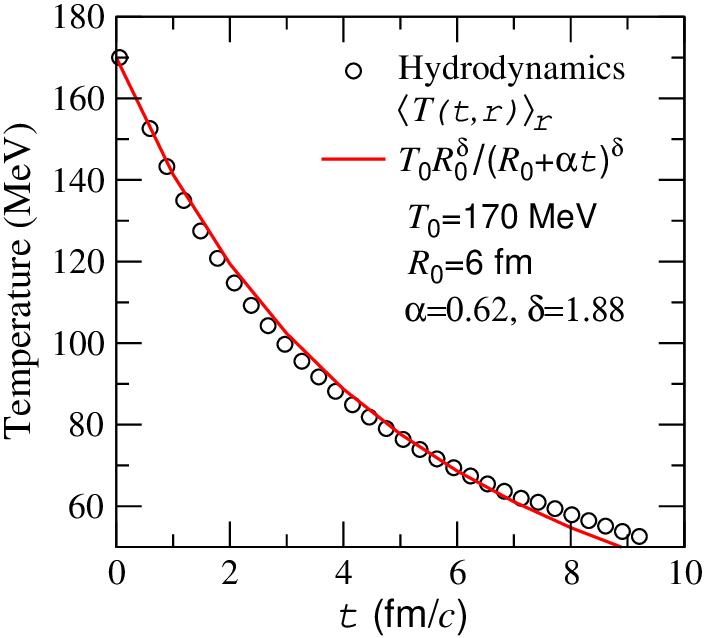}
\caption{(Color online) Source temperature as a function of evolution time.} \label{fTemp}
\vspace*{5mm}
\end{figure}

\begin{figure}
\includegraphics[width=0.4\columnwidth]{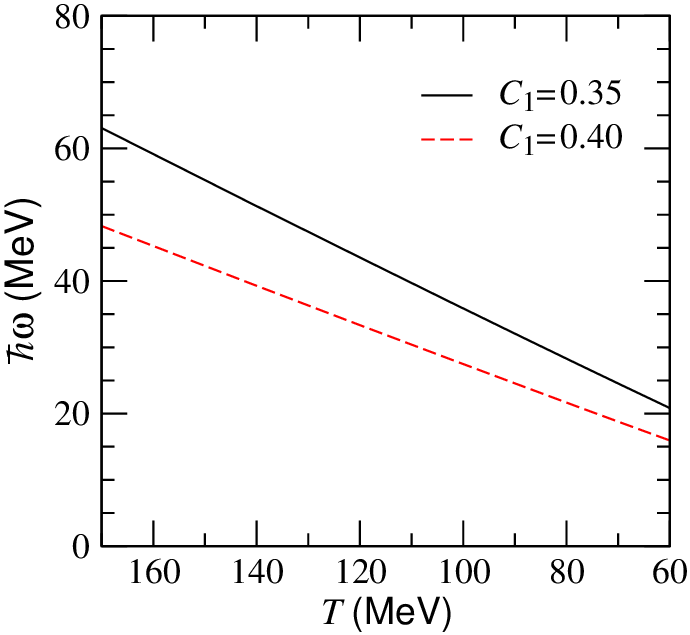}
\caption{(Color online) Potential strength of harmonic oscillator as a function of temperature.}
\label{fhbom}
\end{figure}

Assuming the relaxation time of system is smaller than the source evolution time, we may
approximately deal with the expansion gas as uniform system at each evolution time, as a
first-step improvement to static treatment in the approximation of a quasi-static adiabatic
process.  For such a quasi-static adiabatic expansion of an ideal gas, the system temperature
$T$ and the volume $V$ have the relationship $TV^{\gamma-1}={\rm constant}$, where $\gamma$ is
the ratio of the specific heats at constant pressure and volume.  For example, $\gamma$ is
$5/3$ for non-relativistic monatomic gas.  For our problem, we seek a relation between $T$
and the size of an expanding hadron gas system from relativistic hydrodynamics in high energy
heavy-ion collisions.  We shows in Figure \ref{fTemp} the temperature $T$ averaged over radial
coordinate $r$, $\langle T(t,r)\rangle_r$, obtained from the relativistic hydrodynamic evolution
of a spherical source with an initial radius $R_0=6$ fm and an initial temperature $T_0=170$ MeV.
For the spherical source,  we find that the system temperature satisfies
\begin{equation}
\label{TT}
T=\frac{T_0 R_0^{\delta}}{R^{\delta}}=\frac{T_0 R_0^{\delta}}{(R_0+\alpha t)^{\delta}}\,.
\end{equation}
The fitted line in Fig.\ 1 corresponds to the above formula (\ref{TT}) with $\alpha=0.62$ and
$\delta =1.88$.  We shall use this parameterized formula of the temperature as determined by
relativistic hydrodynamics in our calculations and shall take the characteristic length $a$
in Eq. (\ref{Vrt}) as,
\begin{equation}
\label{Ea}
a=C_1(R_0+\alpha t),
\end{equation}
where $C_1$ is a proportional parameter.  It will be seen that the results of source
root-mean-squared radius are sensitive to the parameter $C_1$.  So, they may provide a
restriction to $C_1$ value.  We plot in Fig. \ref{fhbom} the strength of harmonic oscillator
potential, $\hbar\omega$, as a function of temperature for $C_1=$ 0.35 and 0.40, respectively.
The potential strength decreases with evolution time for the expanding source, and thus
decreases with decreasing temperature.

\subsection{Pion Bose-Einstein condensation}
In relativistic heavy-ion collisions, the source temperature of hadron gas phase is in the region
of [60 to 170] MeV.  At these temperatures pion motion is relativistic.  For a quasi-static process,
at each time stage the system is treated as a static state, and
the eigenvalue equation for relativistic pion with only a scalar interaction as
in Eq. (\ref{Vrt}) is given by \cite{WonZha07}
\begin{equation}
\bigg[\frac{\textbf{\emph{p}}^2}{2m}+V(\textbf{\emph{r}})\bigg]U(\textbf{\emph{r}})=\frac{E^2-m^2}
{2m}U(\textbf{\emph{r}})= \varepsilon U(\textbf{\emph{r}}).
\end{equation}
Here the potential $V(\textbf{\emph{r}})$ is given by Eq. (\ref{Vrt}) at each time stage, for instance
at $t=t_1$, $t_2,\, \cdots,\, t_m$, $V(\textbf{\emph{r}})=\hbar\,\omega(t_1)\,r^2/[2a^2(t_1)],\,
\hbar\,\omega(t_2)\,r^2/[2a^2(t_2)],\,\cdots,\,\hbar\,\omega(t_m)\,r^2/[2a^2(t_m)]$.
The eigenenergy of the relativistic pion is
\begin{equation}
E_n=\sqrt{m^2+2m\varepsilon_n},
\end{equation}
where $\varepsilon_n=(n+3/2)\hbar\omega,~n=0,1,2,\ldots$, and the eigenfunction is
\begin{equation}
\label{wfsn}
U_n(\textbf{\emph{r}})=N_{n_r,\,l} \bigg(\frac{r}{a}\bigg)^l e^{-\frac{r^2}{2a^2}}\,
L_{n_r}^{l+\frac{1}{2}}\bigg(\frac{r^2}{a^2}\bigg) Y_{lm}(\theta,\varphi),
\end{equation}
where, $n = 2n_r+l$,
\begin{equation}
\label{wfsc}
N_{n_r,\,l}=\bigg[\frac{2n!}{a^{3}\Gamma(n_{r}+l+\frac{3}{2})}\bigg]^{1/2},
\end{equation}
$L$ and $Y$ are Laguerre polynomial and spherical harmonics, respectively.  The eigenfunction
of the ground state is
\begin{equation}
\label{wfs0}
U_0(\textbf{\emph{r}})=\bigg(\frac{1}{a^2\pi}\bigg)^{3/4} e^{-\frac{r^2}{2a^2}}\,.
\end{equation}
For a quasi-static expansion, $\omega$ or $a$ is a function of the evolution time.

We introduce $\tilde E_n$ to measure the relative energy levels to the energy of the $n=0$
state,
\begin{equation}
\tilde E_n=E_n-\sqrt{m^2+2m\times \frac{3}{2}\hbar\,\omega}.
\end{equation}
For the identical boson gas with a fixed number of particles, $N$, and at a given temperature
$T=1/\beta$, we have
\begin{equation}
\label{N0T}
N=N_0+N_T,
\end{equation}
where, $N_0$ is the number of condensate particles in $n=0$ state,
\begin{equation}
\label{N0}
N_0=\frac{\mathcal Z}{1-\mathcal Z},
\end{equation}
$N_T$ is the number of the particles in $n>0$ states,
\begin{equation}
\label{NTn}
N_T=\sum_{n>0}^{\infty}\frac{g_n \mathcal Z\,e^{-\beta\tilde E_n}}
{1-\mathcal Z\,e^{-\beta\tilde E_n}},
\end{equation}
and $\mathcal Z$ is the fugacity parameter which includes the factor for the lowest energy
$\varepsilon_0$ \cite{Nar99,WonZha07}.  Because $N_0 \ge 0$, the values of $\mathcal Z$ are
between zero and one.  When temperature is lowered below the critical temperature $T_{\rm c}$,
Bose-Einstein condensation occurs.  In this case, $N_0 \sim N$ and $\mathcal Z \sim N/(N+1)$.

\begin{figure}
\includegraphics[width=0.77\columnwidth]{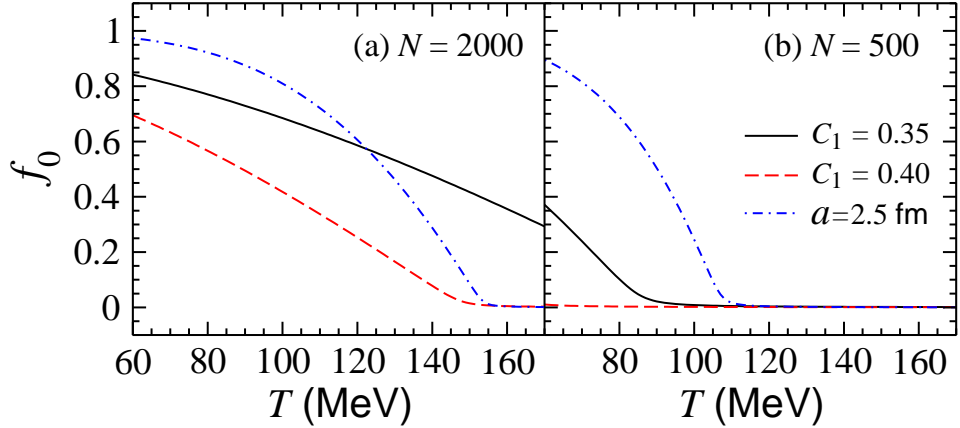}
\caption{(Color online) Source condensation fraction as a function of temperature.}
\label{zf0}
\end{figure}

From Eqs. (\ref{N0T}), (\ref{N0}), and (\ref{NTn}), one can calculate the fugacity parameter
$\mathcal Z$ for fixed $N$ numerically, and then obtain the condensation fraction
\cite{WonZha07,LiuZha13},
\begin{equation}
f_0=\frac{N_0}{N}=\frac{\mathcal Z}{(1-\mathcal Z)N}\,.
\end{equation}
In Figs. \ref{zf0}(a) and \ref{zf0}(b), we show the condensation fraction for the sources with
$N=$ 2000 and 500, respectively.  Here, the solid and dashed lines are for the proportional
parameter $C_1$ in Eq.\ (\ref{Ea}) being taken as 0.35 and 0.40, and the dashed-dot lines are
for the static source with a fixed characteristic length $a=2.5$ fm ($R=R_0=6$ fm, $C_1=0.417$)
for comparison.  One can see that the condensation fraction $f_0$ decreases with increasing
temperature.  In Fig. \ref{zf0}(a), the values of $f_0$ for the $C_1=0.40$ case and for the
fixed $a$ case both approach zero at $T\sim 160$ MeV, corresponding to the completely uncondensed 
case.  However, the condensation fraction for $C_1=0.35$ is finite at the initial high
temperature for the source with $N=2000$.  For the expanding sources $f_0$ increases more
slowly with decreasing temperature than that for the static source.  From Fig. \ref{zf0}(b)
it can be seen that the expanding source with $N=500$ and $C_1=0.35$ has only a small
condensation fraction at low temperatures, and the system with $C_1=0.40$ is uncondensed
in the whole temperature region.

\section{Density distributions and correlation functions}
Based on quantum statistics, the spatial one-body density matrix for the boson gas with given
particle number and temperature is
\begin{equation}
\label{G1r}
G^{(1)}(\textbf{\emph{r}}_1,\textbf{\emph{r}}_2)=\sum_{n=0}^{\infty} U_n^*(\textbf{\emph{r}}_1)
U_n(\textbf{\emph{r}}_2) \frac{g_n\mathcal Z\,e^{-\beta\tilde E_n}}{1-\mathcal Z\,e^{-\beta\tilde
E_n}}\,,
\end{equation}
and the density matrix in momentum space is
\begin{equation}
\label{G1p}
G^{(1)}(\textbf{\emph{p}}_1,\textbf{\emph{p}}_2)=\sum_{n=0}^{\infty} U_n^*(\textbf{\emph{p}}_1)
U_n(\textbf{\emph{p}}_2) \frac{g_n\mathcal Z\,e^{-\beta\tilde E_n}}{1-\mathcal Z\,e^{-\beta\tilde
E_n}}\,,
\end{equation}
where $U_n(\textbf{\emph{p}})$ are the eigenfunctions in momentum space, which have the exchange
symmetry $\textbf{\emph{p}}a/\hbar$ --- $\textbf{\emph{r}}/a$ with the spatial eigenfunctions
$U_n(\textbf{\emph{r}})$ [Eqs. (\ref{wfsn}) and (\ref{wfs0})] for harmonic oscillator potential.

In the limit of a large number of particles, $N(N-1)\sim N^2(\,\gg N_T,N_0)$, the two-body
density matrices can be written as \cite{Nar99,Pol96,WonZha07}
\begin{eqnarray}
\label{G2r}
G^{(2)}(\textbf{\emph{r}}_1,\textbf{\emph{r}}_2;\textbf{\emph{r}}_1,\textbf{\emph{r}}_2)
&=&G^{(1)}(\textbf{\emph{r}}_1,\textbf{\emph{r}}_1)\,G^{(1)}(\textbf{\emph{r}}_2,
\textbf{\emph{r}}_2) +|G^{(1)}(\textbf{\emph{r}}_1,\textbf{\emph{r}}_2)|^2\cr
&&-N_0^2|U_0(\textbf{\emph{r}}_1)|^2|U_0(\textbf{\emph{r}}_2)|^2\,,
\end{eqnarray}
\begin{eqnarray}
\label{G2p}
G^{(2)}(\textbf{\emph{p}}_1,\textbf{\emph{p}}_2;\textbf{\emph{p}}_1,\textbf{\emph{p}}_2)
&=&G^{(1)}(\textbf{\emph{p}}_1,\textbf{\emph{p}}_1)G^{(1)}\,(\textbf{\emph{p}}_2,
\textbf{\emph{p}}_2) +|G^{(1)}(\textbf{\emph{p}}_1,\textbf{\emph{p}}_2)|^2\cr
&&-N_0^2|U_0(\textbf{\emph{p}}_1)|^2|U_0(\textbf{\emph{p}}_2)|^2\,.
\end{eqnarray}

\subsection{Density distributions}
With density matrices, we can obtain the density distributions in configuration and
momentum spaces as
\begin{equation}
\label{rhor}
\rho(\textbf{\emph{r}})=G^{(1)}(\textbf{\emph{r}},\textbf{\emph{r}})=\sum_{n=0}^{\infty}
\frac{g_n \mathcal Z\,e^{-\beta\tilde E_n}}{1-\mathcal Z\,e^{-\beta\tilde E_n}}
|U_n(\textbf{\emph{r}})|^2,
\end{equation}
\begin{equation}
\label{rhop}
\rho(\textbf{\emph{p}})=G^{(1)}(\textbf{\emph{p}},\textbf{\emph{p}})=\sum_{n=0}^{\infty}
\frac{g_n \mathcal Z\,e^{-\beta\tilde E_n}}{1-\mathcal Z\,e^{-\beta\tilde E_n}}
|U_n(\textbf{\emph{p}})|^2.
\end{equation}

In Figs. \ref{rhor}(a) and \ref{rhor}(b) we plot the spatial density distributions of the
expanding sources with $C_1=$ 0.35 and 0.40, and at different temperatures, respectively.
In Fig. \ref{rhor}(c) we plot the spatial density distributions of the static source with
$a=2.5$ fm for comparison.  The particle number of the systems is 2000.  At $T=160$ MeV,
the distribution for $C_1=0.35$ has an obvious rise in the small $r$ region as compared
to those for $C_1=0.40$ and $a=2.5$ fm.  This is because the source with $C_1=0.35$ has
a finite degree of Bose-Einstein condensation at $T=160$ MeV, and the other systems are
completely uncondensed at this temperature (see Fig. \ref{zf0}).  Similarly, one can see
the two-tiered structure of the density distributions caused by the condensation for all
the three systems at the lower temperatures.  Because the source with $C_1=0.40$ has lower
condensation fraction at the lower temperatures, its corresponding density distributions
are wider than those of the other systems.

\begin{figure}
\includegraphics[width=0.62\columnwidth]{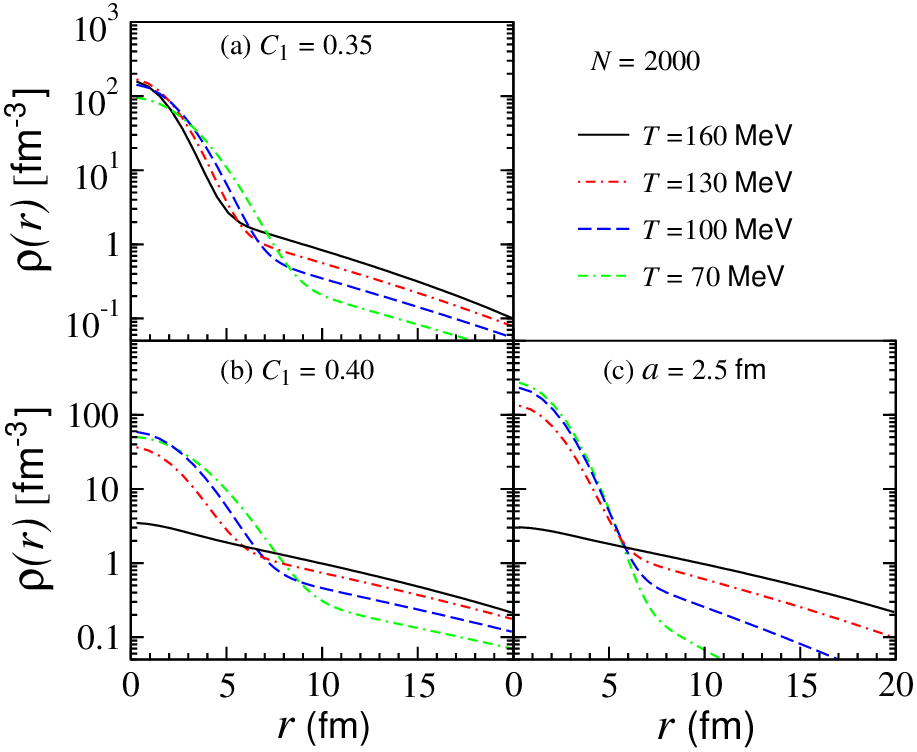}
\caption{(Color online) Spatial density distributions of the expanding sources with $C_1=$
0.35 and 0.40, and the system with fixed $a=2.5$ fm.  The particle number of the systems
is 2000. }
\label{rhor}
\end{figure}

\begin{figure}
\vspace*{5mm}
\includegraphics[width=0.62\columnwidth]{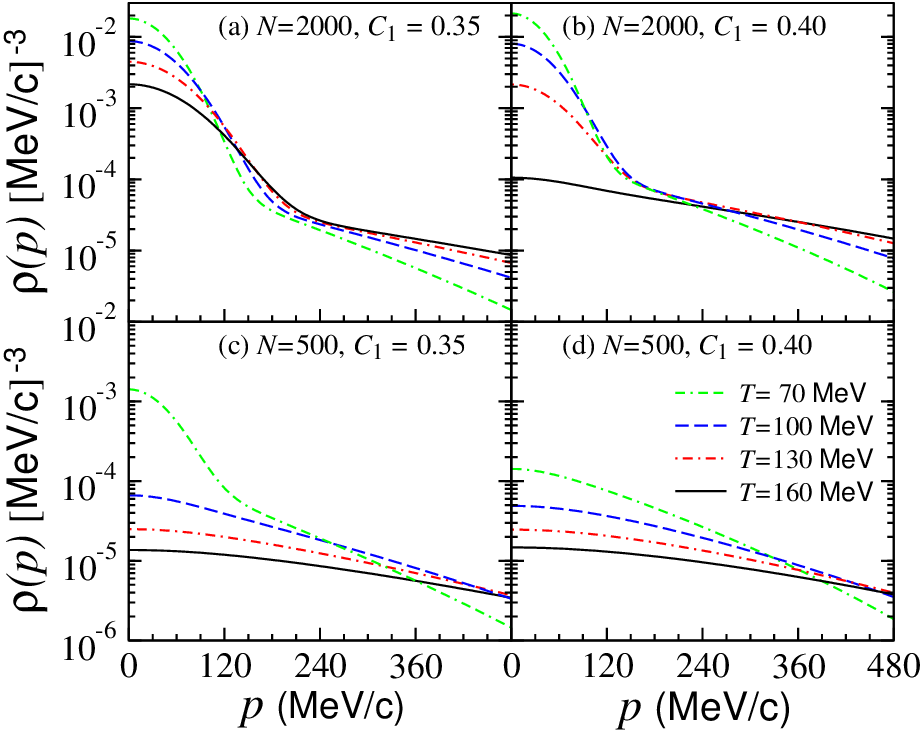}
\caption{(Color online) The density distributions in momentum space for the expanding sources
with $N=$ 2000 and 500, and at different temperatures. }
\label{rhop}
\end{figure}

In Fig. \ref{rhop}, we plot the density distributions in momentum space for the expanding
sources with $N=$ 2000 and 500.  One can also see the two-tiered structure in the momentum
density distributions in Figs. \ref{rhop}(a), \ref{rhop}(b), and \ref{rhop}(c), arising from
the Bose-Einstein condensation at lower temperatures.  The average particle momentum is lower
for the source with a higher degree of condensation.  For the completely uncondensed source
in Fig. \ref{rhop}(d), there is no two-tiered structure of momentum density distribution in
the whole temperature region.  The momentum density distribution becomes wider when the
temperature increases.

\subsection{Two-particle momentum correlation functions}
Using density matrices, the two-particle correlation function in momentum space can be written
as
\begin{equation}
\label{Cp1p2_1}
C(\textbf{\emph{p}}_1,\textbf{\emph{p}}_2)=\frac{G^{(2)}(\textbf{\emph{p}}_{1},{\bf p}_2;
\textbf{\emph{p}}_{\!1},\textbf{\emph{p}}_2)}{G^{(1)}(\textbf{\emph{p}}_1,\textbf{\emph{p}}_1)
\,G^{(1)}(\textbf{\emph{p}}_2,\textbf{\emph{p}}_2)}.
\end{equation}
From Eq. (\ref{G2p}), we have
\begin{equation}
\label{Cp1p2_2}
C(\textbf{\emph{p}}_1,\textbf{\emph{p}}_2)=1 +\frac{|G^{(1)}(\textbf{\emph{p}}_1,
\textbf{\emph{p}}_2)|^2 -N_0^2|U_0(\textbf{\emph{p}}_1)|^2 |U_0(\textbf{\emph{p}}_2)|^2}
{G^{1}(\textbf{\emph{p}}_1,\textbf{\emph{p}}_1)\, G^{(1)}(\textbf{\emph{p}}_2,
\textbf{\emph{p}}_2)}.
\end{equation}
In the nearly completely coherent case with almost all particles in the ground condensate
state, $N_0 \to N$, the two terms in the numerator cancel each other and we have
$C(\textbf{\emph{p}}_1,\textbf{\emph{p}}_2)=1$, as it should be.  For the other extreme
of a completely chaotic system with $N_0 \ll N$, the second term in the numerator can be
neglected and we have the usual one for a completely chaotic source,
\begin{equation}
\label{Ccha}
C(\textbf{\emph{p}}_1,\textbf{\emph{p}}_2)=1+\frac{|G^{(1)}(\textbf{\emph{p}}_1,
\textbf{\emph{p}}_2)|^2} {G^{(1)}(\textbf{\emph{p}}_1,\textbf{\emph{p}}_1)\,
G^{(1)}(\textbf{\emph{p}}_2, \textbf{\emph{p}}_2)}.
\end{equation}

It is convenient in interferometry analyses to introduce the average and relative momenta
\begin{equation}
\textbf{\emph{p}}=(\textbf{\emph{p}}_1 +\textbf{\emph{p}}_2)/2, ~~~~~~
\textbf{\emph{q}}=\textbf{\emph{p}}_1 -\textbf{\emph{p}}_2.
\end{equation}
The momentum correlation function can be expressed alternatively in terms of the kinematic
variables, $\textbf{\emph{p}}$ and $\textbf{\emph{q}}$, and the correlation function $C(p,q)$
can be obtained in numerical calculations by integrating $\textbf{\emph{p}}_1$ and
$\textbf{\emph{p}}_2$ in Eq. (\ref{Cp1p2_2}) for certain $(p,\,q)$ bins.

\begin{figure}
\includegraphics[width=0.6\columnwidth]{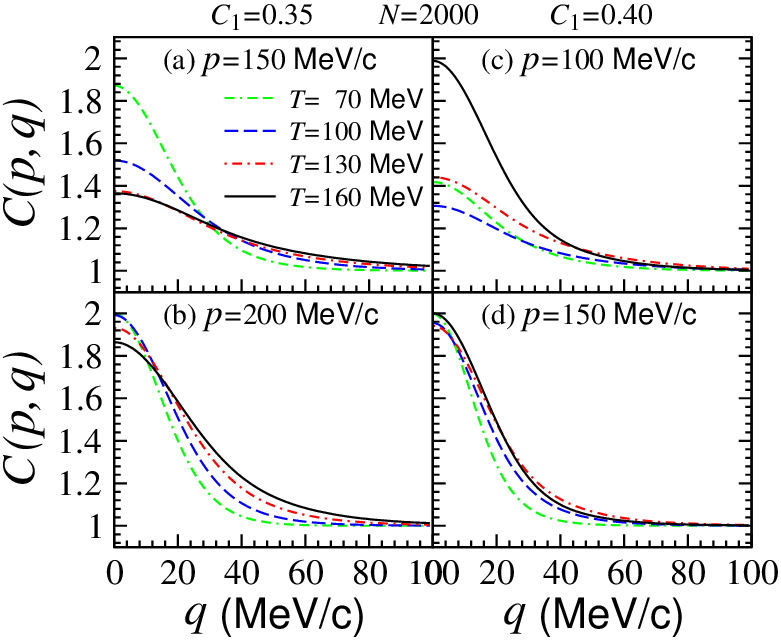}
\caption{(Color online) The two-pion correlation functions $C(p,q)$ for the sources with
$N=2000$, $C_1=$ 0.35 and 0.40. }
\label{cpq2000}
\end{figure}

In Fig. \ref{cpq2000} we plot the two-pion correlation functions $C(p,q)$  for the sources
with $N=2000$, $C_1=$ 0.35 and 0.40.  One can see that at the higher momenta, $p=200$ Mev
for $C_1=0.35$ [Fig. \ref{cpq2000}(b)] and $p=150$ MeV for $C_1=0.40$ [Fig. \ref{cpq2000}(d)],
the width of the correlation functions decrease when the  temperature decreases.  This is
because the source expansion leads to a decrease temperature.  The intercepts of the
correlation functions at the higher momenta are larger than the corresponding results at
the lower momenta [Figs. \ref{cpq2000}(a) and \ref{cpq2000}(c)] because the particles with
large momenta are from the uncondensed chaotic states on the average, for the finite
condensation sources.  One can also see from Figs. \ref{cpq2000}(a) and \ref{cpq2000}(c)
that at the lower momenta the intercept varies with temperature in a non-monotonic manner.
It will be discussed in detail in next section that the complexity of  this variation arises
because the intercept value depends not only  on the condensation fraction but also on
the radius of expanding source,  both  varying  with temperature.

In actual interferometry analyses, the source expansion may lead to a space-momentum
correlation so that the two particles with small relative momentum are from neighboring
source points \cite{Heinz02,Lis05,Zhang12}.  This space-momentum correlation can influence
the construction of HBT correlation functions and therefore the dependence of HBT radius
as a function of particle pair momentum \cite{Heinz02,Lis05,Zhang12}.  Note that the HBT
correlation functions calculated by the density matrices has not included the effect of
the space-momentum correlation.  So, it is hard to compare the HBT radii obtained from
the correlation functions calculated here with those from the actual interferometry
analyses for expanding sources.  However, the effect of Bose-Einstein condensation on
the chaoticity parameter $\lambda$ in HBT interferometry should be independent of this
space-momentum correlation in principle, because the source condensation depends only
on the source thermal environment.

\section{Chaoticity parameter $\lambda$ in two-pion interferometry}
In HBT analyses the chaoticity parameter $\lambda$ is introduced phenomenologically to
represent the intercept of the HBT correlation function at zero relative momenta of the
particle pair,
\begin{equation}
\lambda(\textbf{\emph{p}})=C(\,\textbf{\emph{p}},\textbf{\emph{q}}=0)-1\,.
\end{equation}
From Eq. (\ref{Cp1p2_2}), the chaoticity parameter can be expressed in terms of $N_0=f_0N$
and the ratio of the squared of ground state wave function $U_0(\textbf{\emph{p}})$ to the
momentum density $\rho(\textbf{\emph{p}})=G^{(1)}(\textbf{\emph{p}},\textbf{\emph{p}})$,
\begin{equation}
\label{lambda}
\lambda(\textbf{\emph{p}})=1-N_0^2\big[\,|U_0(\textbf{\emph{p}})|^2/\rho(\textbf{\emph{p}})
\big]^2 \equiv 1-\big[f_0 F_N(\textbf{\emph{p}})\big]^2\,,
\end{equation}
where
\begin{equation}
F_N(\textbf{\emph{p}})=N |U_0(\textbf{\emph{p}})|^2/\rho(\textbf{\emph{p}})\,.
\end{equation}
$F_N(p)$ gives the relative probability of particle pairs from the condensed state with
momentum $p$ to the momentum density.  For the source with a finite condensation, $F_N(p)$
at large $p$ will decrease as the momentum $p$ increases because $|U_0(p)|^2$ decrease more
rapidly than $\rho(p)$ as $p$ increases.

\begin{figure}[tbp]
\includegraphics[width=0.6\columnwidth]{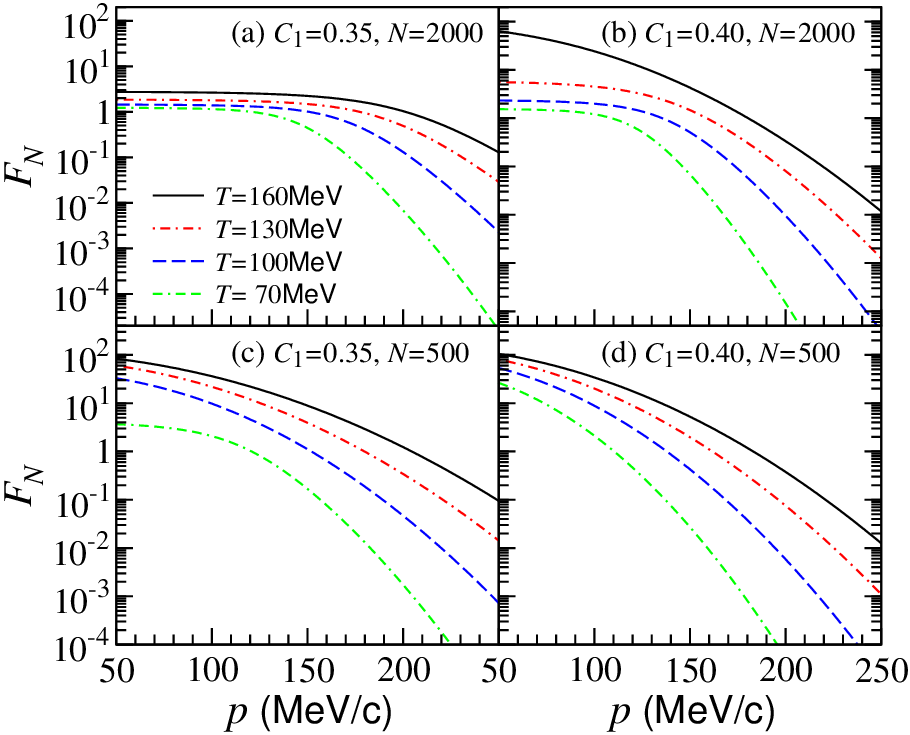}
\caption{(Color online) $F_N$ as a function of momentum for the systems with $N=$ 2000 and
500, $C_1=$ 0.35 and 4.0, and at different temperatures. }
\label{u0rho_p}
\end{figure}

In Fig. \ref{u0rho_p}, we plot $F_N$ as a function of momentum at different temperatures for
the sources with $N=$ 2000 and 500.  From Fig. \ref{u0rho_p}(a) one can see that the variation
of $F_N$ with momentum has an obvious abruptly changing point.  For example, the curve of
$F_N$ for $T=160$ MeV in Fig. \ref{u0rho_p}(a) decreases more rapidly at $p\sim$ 200 MeV/$c$.
The reason is that the density distribution $\rho(p)$ changes abruptly as a function of
momentum at $p\sim$ 200 MeV/$c$ [see Fig. \ref{rhop}(a)] for the source with a finite
condensation fraction $f_0$ [see Fig. \ref{zf0}(a)].  With decreasing temperature, $f_0$
increases, and the changing point of momentum of $F_N(p)$ decreases.  The values of $F_N$ at
small momentum decrease with decreasing temperature.  This is because when the temperature
decreases the characteristic length $a$ increases, leading to the decrease of $|U_0(p)|^2$,
and $\rho(p)$ is higher at small momentum for the lower temperature.  For the completely
uncondensed cases, for example, for the $T=160$ MeV case in Fig. \ref{u0rho_p}(b),  the $T=$
160, 130, and 100 MeV cases in Fig. \ref{u0rho_p}(c), and all cases in Fig. \ref{u0rho_p}(d),
there are no abruptly change points in $F_N$ as the momentum increases.  The values of $F_N$
increase with temperature because the distribution of $\rho(p)$ becomes wider for the source
with a smaller radius when the system is at higher temperature.

\begin{figure}[tbp]
\includegraphics[width=0.7\columnwidth]{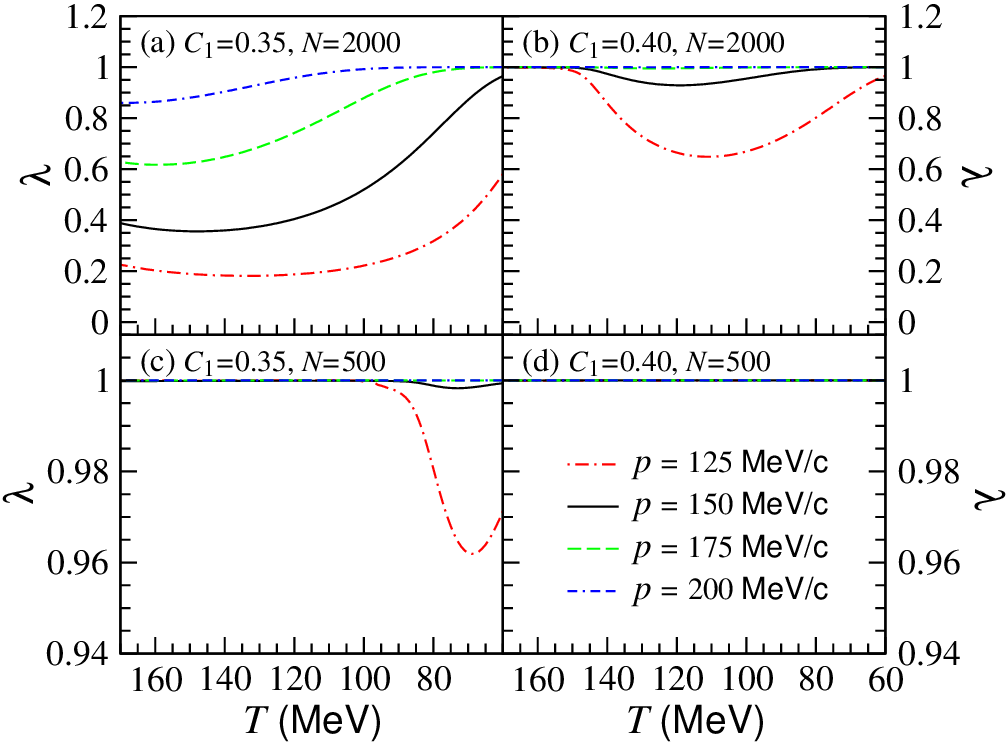}
\caption{(Color online) The chaoticity parameter $\lambda$ as a function of temperature for
different pion pair momentum values. }
\label{lamb_T}
\end{figure}

\begin{figure}[tbp]
\vspace*{5mm}
\includegraphics[width=0.85\columnwidth]{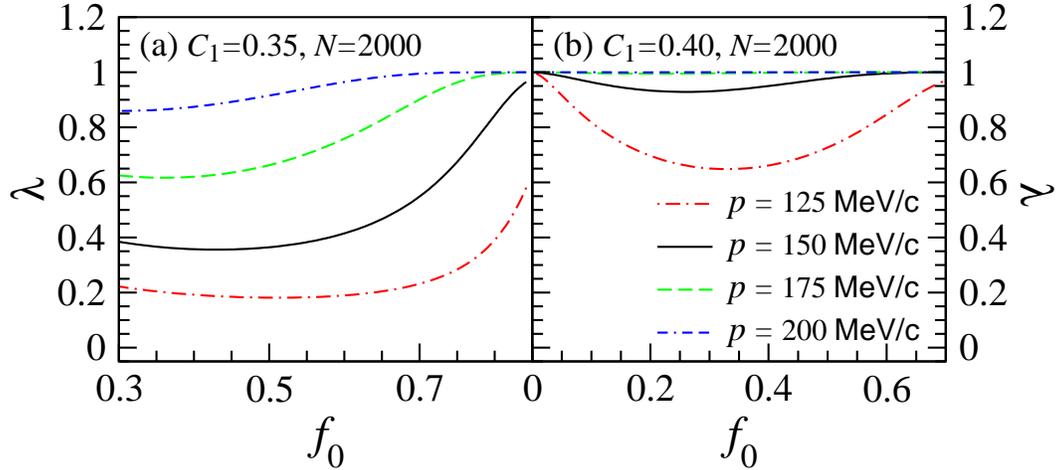}
\caption{(Color online) The chaoticity parameter $\lambda$ as a function of $f_0$ for different
pion pair momentum values and $N=2000$. }
\label{lamb_f0}
\end{figure}

In Fig. \ref{lamb_T}, we plot the chaoticity parameter $\lambda$ as a function of temperature
$T$ for different pair momentum values.  For the source with $N=2000$ and $C_1=0.35$, at $T=
170$ MeV in Fig. \ref{lamb_T}(a), the $\lambda$ values are less than one and they increase with
increasing  momentum $p$.  This is because that the source has a finite but small condensation
fraction at that temperature, and the particles with large momenta are from the higher-energy
uncondensed states on the average.  When the temperature decreases, the condensation fraction
$f_0$ increases but $F_N$ decreases especially at large momentum.  The competition of $f_0$
and $F_N$ results in the value of $\lambda$ as a function of temperature.
In Fig. \ref{lamb_T}(b), the $\lambda$ value is one at temperatures greater than 150 MeV
because the source is completely uncondensed at these temperatures, and thus $f_0=0$.
For the smaller particle number source in Figs. 8(c) and 8(d), one can see that the
$\lambda$ results deviate from one only for small momenta and at low temperatures.  The
chaoticity parameter results in Fig. \ref{lamb_T}(a) and Fig. \ref{lamb_T}(b) are consistent
with the intercepts of the two-pion correlation functions shown in Fig. \ref{cpq2000}.
Figures \ref{lamb_f0}(a) and \ref{lamb_f0}(b) show the variation of $\lambda$ with the
condensation fraction $f_0$ for the source with $N=2000$, $C_1=$ 0.35 and 0.40, respectively.
The appearances of the curves are similar to those shown in Figs. \ref{lamb_T}(a) and
\ref{lamb_T}(b) because $f_0$ varies almost linearly with temperature for the expanding
sources [see the solid and dashed lines in Fig. \ref{zf0}(a)].  The variation of $\lambda$
with $f_0$ is different from the results of the static source where $\lambda$ decreases
with $f_0$ monotonically (see the Fig. 11 in Ref. \cite{LiuZha13}).

\begin{figure}
\vspace*{0mm}
\includegraphics[width=0.85\columnwidth]{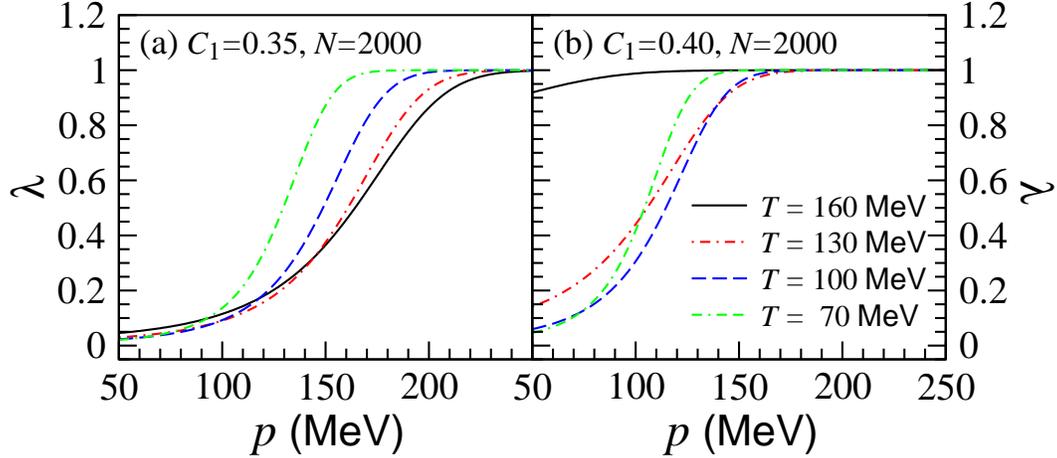}
\caption{(Color online) The chaotic parameter $\lambda$ as a function of pion pair momentum
at different temperatures. }
\label{lamb_p}
\end{figure}

In Fig. \ref{lamb_p}, we plot the chaotic parameter $\lambda$ as a function of the momentum
at different temperatures for the sources with $N=2000$.  The values of $\lambda$ increase
with increasing momentum and approach to one at high momenta because the particles with large
momenta are from the uncondensed high-energy states on the average, for the sources with a
finite and small condensation fraction.  For an expanding source, $U_0(p)$ is also a function
of the temperature because the characteristic length $a$ increases when the temperature
decreases.  So, the increases of $\lambda$ with the momentum at different temperatures
exhibit more complexity than the results for the static source (see the Fig. 16 in Ref.
\cite{LiuZha13}).

\begin{figure}[tbp]
\vspace*{5mm}
\includegraphics[width=0.5\columnwidth]{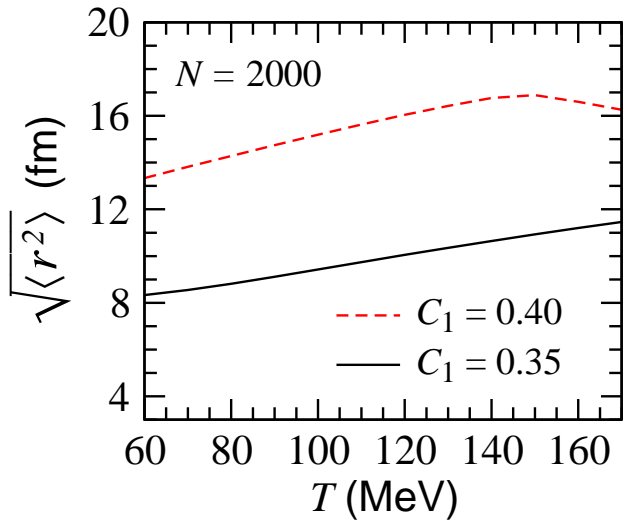}
\caption{(Color online) The root-mean-squared radius of the sources with $N=2000$, $C_1=$
0.35 and 0.40. }
\label{rmsrT}
\end{figure}

In the heavy-ion collisions at the LHC energy, the identical pion multiplicity of event can
reach several thousands.  The calculations indicate that in this case the effect of
Bose-Einstein condensation on the chaotic parameter $\lambda$ in two-pion HBT interferometry
may be considerable and should be taken into account.  In Fig. \ref{rmsrT} we plot the
root-mean-squared radius (RMSR) of the source with $N=2000$, $C_1=$ 0.35 and 4.0.  It can
be seen that the results of the RMSR are much sensitive to the parameter $C_1$.  Considering
the RMSR of the sources in Pb-Pb collisions at the LHC to be of the order 10 fm, the values
of parameter $C_1$ between 0.35 and 0.40 appear reasonable.

\section{Summary and conclusion}
We investigate pion gas Bose-Einstein condensation in relativistic heavy-ion collisions in
an expanding source model.  The relationship between the source temperature and radius is
given by a parameterized formula obtained with relativistic hydrodynamic calculations.
Using the one- and two-body density matrices for the boson gas within the mean-field with
a harmonic oscillator potential, we calculate the space and momentum density distributions,
two-pion HBT correlation function, and the chaoticity parameter in two-pion interferometry
for the identical pion sources at the temperatures of the hadronic phase in relativistic
heavy-ion collisions.  The influences of the source particle number and the potential
strength of mean-field on the density distributions, HBT correlation functions, and $\lambda$
values are discussed.

In the heavy-ion collisions at the LHC energy, the identical pion multiplicity of event can
reach several thousands.  Our investigations indicate that the sources with thousands of
identical pions may exhibit a degree of Bose-Einstein condensation at the temperatures of the
hadronic phase, 170-- 60 MeV, in relativistic heavy-ion collisions.  This finite condensation
may decrease the chaoticity parameter $\lambda$ in the two-pion interferometry measurements at
low pion pair momenta, and influence very slightly the $\lambda$ value at high pion pair
momentum.  Unlike the results of static source, the chaoticity parameter $\lambda$ for the
expanding source depends not just on the condensation fraction $f_0$ alone.  Its variation
with the temperature is more complicated.  In experiments, the source temperature can be
measured by the slope of momentum spectrum.  Once the temperature is determined, it is of
interest to compare the model $\lambda$ values with experimental HBT data at different
momenta.

In this paper we investigate the influence of Bose-Einstein condensation on the HBT Chaoticity
parameter.  There are many other effects, such as Coulomb interaction, that may influence
the chaoticity parameter measurements in experiments.  Further investigations of the other
effects on $\lambda$ measurements to separate out the effect of Bose-Einstein condensation
on pion momentum spectrum and HBT measurements in two-pion and multi-pion interferometry will
be of great interest.

\begin{acknowledgments}
This research was supported by the National Natural Science Foundation of China under Grant
No. 11275037.
\end{acknowledgments}

\end{document}